\documentclass[aps,prbbib,floatfix,floats]{revtex4}
\usepackage{graphicx,latexsym}
\usepackage{epsfig}
\usepackage{amsmath}
\usepackage{color}
\begin{document}
\title{\bf Transport through quantum dots with magnetic impurities}

\author{M. \c{T}olea and A. Aldea}
\affiliation{National Institute of Materials Physics, POBox MG7,
Bucharest-Magurele, Romania}
\author{B. R. Bu{\l}ka}
\affiliation{Institute of Molecular Physics, Polish Academy of
Science, ul. M. Smoluchowskiego 17, 60-179 Pozna{\'n}, Poland}

\begin{abstract}

We analyze the electronic transport through a quantum dot that
contains a magnetic impurity. The coherent transport of electrons is
governed by the quantum confinement inside the dot, but is also
influenced by the exchange interaction with the impurity. The
interplay between the two gives raise to the singlet-triplet
splitting of the energy levels available for the tunneling electron.
In this paper, we focus on the charge fluctuations and, more
precisely, the height of the conductance peaks. We show that the
conductance peaks corresponding to the triplet levels are three
times higher than those corresponding to singlet levels, if
electronic correlations are neglected (for non-interacting dots, when an exact
solution can be obtained).
Next, we consider the Coulomb repulsion and the many-body
correlations. In this case, the singlet/triplet peak height ratio
has a complex behavior. Usually the highest peak corresponds to the
state that is lowest in energy (ground state), regardless if it is
singlet or triplet. In the end, we get an insight on the Kondo
regime for such a system, and show the formation of three Kondo
peaks. We use the equation of motion method with appropriate
decoupling.


\end{abstract}
\maketitle

\section{Introduction}
The quantum dots are zero-dimensional structures (so called
artificial atoms) that are possible to be experimentally obtained
and investigated individually, for some years now. An incoming
electron may tunnel through a quantum dot only if its energy matches
a dot energy level, so we have a single electron transistor
\cite{Kastner}. Another interesting potential application is quantum
computing using the spin of the quantum dot as information carrier
\cite{Loss}. The parameters that influence transmittance through a
quantum dot (like the size of the dot or the hybridization with the
leads) are much more easy to control comparing to the parameters of
a bulk system, so one can test quantum theories in a way that was
not possible before. We give as example the observation of the
two-channel Kondo effect, described theoretically many years ago \cite{2channel1,2channel2},
but realized experimentally only very recently, in a double-dot
system \cite{Potok} .

In the last years, people have started to address the problem of
quantum dots with magnetic impurities
\cite{Govorov,Murthy,Aldea,Heersche,Kaul,Rossier,TB}, which is also
the subject of the present paper. The transient electron feels the
exchange interaction with the impurity and forms singlet/triplet
entangled states before tunneling out of the dot. We analyze the way
in which this physics is seen in transport. It will be shown that
the peaks corresponding to the triplet levels are three times higher
than those corresponding to singlet levels, if we neglect the
Coulomb repulsion and electronic correlations in the dot. In large dots, with low confinement,
the Coulomb repulsion is reduced, and the approximation to neglect it may
still capture some important physics.
Such a case will be addressed in Section II.

The case of small quantum dots, with strong Coulomb repulsion is treated in
Section III.
If the Coulomb interaction and the
many-body correlations are considered, the picture becomes more
complex: the peak corresponding to the ground state is usually
higher, regardless if it is singlet or triplet. The peaks height
depends on the ratio $J/\Gamma$ (exchange interaction versus coupling strength with the leads).
An analytical formula is proposed for the
transmittance and in particular for the singlet/triplet peak height
ratio. In the end, we get an insight on the Kondo regime and show
the formation of three Kondo peaks in the density of states.

\section{Large quantum dots with magnetic impurities}

In this section we shall neglect all electron-electron interactions,
except for the interaction with the magnetic impurity. The correlations
are neglected as well, and they will be considered in the next
section. The problem presented in this section is a two-electron
scattering problem (one of which -the impurity- is fixed and may
only change its spin orientation).

If correlations are neglected, one can easily afford to consider a
many-site dot, which is realistic from geometrical point of view. In
(large) quantum dots with low confinement, the Coulomb repulsion is
reduced, and this justifies the non-interacting model used here. A
general lattice Hamiltonian, with a localized spin interaction, can
be written as:
\begin{eqnarray}
H_{S}&=& \sum_{\langle i,j\rangle,\sigma}w_{ij,\sigma}~ e^{i2\pi
\Phi_{ij}}c_{i,\sigma}^\dagger c_{j,\sigma}  +
\sum_{\sigma,i\in QD}V_g c_{i,\sigma}^{\dagger} c_{i,\sigma}\nonumber\\
&+&\frac12J~ (c^\dagger_{n\uparrow} c_{n\uparrow}-
c^\dagger_{n\downarrow} c_{n\downarrow})S^z_n
+\frac12J~(c^\dagger_{n\uparrow} c_{n\downarrow}S^-_n+c^\dagger_{n\downarrow} c_{n\uparrow}S^+_n)~;
~~~~n\in QD~,
\end{eqnarray}
where ~$c_i^\dagger~ (c_i)$ are creation (annihilation) operators in
the dot sites indexed by $i$~; the index $n$ is devoted to the site
of the dot where the magnetic impurity is placed. $w_{ij,\sigma}$
are hopping parameters (actually the hopping parameters are spin-independent and
in the following we give up the spin index) and $ \Phi_{ij}$ are the phases associated
with an (eventually applied) magnetic field. The last two term account for the interaction
with the magnetic impurity \cite{explain}.

From this point on, we take S=1/2.
 One can define the  fermion operators $\{d^{\dagger}_\uparrow,
d^{\dagger}_\downarrow, d_\uparrow,d_\downarrow\}$ for the localized
spin (provided one projects out all the states with occupancy
different from $1$)

Then, $S^z_n,~ S^+_n and~ S^-_n$ in eq (1) can be written as:
\begin{equation}
%eq11
S^z_n={1\over2}(d^{\dagger}_\uparrow d_\uparrow
-d^{\dagger}_\downarrow d_\downarrow),~~~
S^+_n=d^{\dagger}_{\uparrow} d_{\downarrow},~~
S^-_n=d^{\dagger}_{\downarrow} d_{\uparrow}~.
\end{equation}
 The natural formulation of the
problem is in terms of singlet-triplet operators (see for instance
\cite{Ho}). Let us introduce the singlet operator $\Sigma_i$ and the
triplet operators $T_i^p$ (p=1,2,3):
\begin{eqnarray}
%eq12
\Sigma_i&=&{1\over\sqrt2}~(d_{\uparrow} c_{i{\downarrow}}-
d_{\downarrow} c_{i{\uparrow}}),~ T_i^1={1\over\sqrt2}~(d_{\uparrow}
c_{i{\downarrow}}+d_{\downarrow}
c_{i{\uparrow}}),\nonumber \\
T_i^2&=&d_{\uparrow} c_{i{\uparrow}},~ T_i^3=d_{\downarrow}
c_{i{\downarrow}}~.
\end{eqnarray}
The above operators may be used in order to write the Hamiltonian
(Eq.1) in the following way (where we have used also
$n_d=d^{\dagger}_\uparrow d_\uparrow + d^{\dagger}_\downarrow
d_\downarrow =1$):
%eq13
\begin{eqnarray}
H_{S}= \sum_{\langle i,j\rangle}~(w^{\Sigma}_{ij}~\Sigma_{i}^\dagger
\Sigma_{j}+w^T_{ij}~\sum_{p=1}^{3}~T^{p\dagger}_{i}T^p_j),\nonumber\\
w^{\Sigma}_{ij}=w_{ij}~ e^{i2\pi\Phi_{ij}}+\delta_{ij}V_g-{3\over4 }J
\delta_{ij}
\delta_{in},~\nonumber \\
w^T_{ij}=w_{ij}~ e^{i2\pi\Phi_{ij}}+\delta_{ij}V_g+{1\over4}J
\delta_{ij} \delta_{in}~.
\end{eqnarray}

In order to analyze transport properties, one needs to connect leads
to the quantum dot. The leads are modeled by a one-dimensional chain
, consistent with the dot tight-binding model.
For the non-interacting dot (more precisely: the interaction is only with
the magnetic impurity, and the electron-electron interactions and correlations
are neglected),
the introduction of
leads is equivalent to the introduction of a complex selfenergy in
the Hamiltonian, at the contact sites. The procedure is described in
detail in \cite{Aldea}. It results the effective Hamiltonian:

\begin{eqnarray}
H^{eff}_{S}&=& H_{S} + \sum_{\alpha,\sigma}\tau_\alpha^2 e^{-iq}~
c_{\alpha,\sigma}^\dagger c_{\alpha,\sigma}\nonumber\\
&=&\sum_{\langle i,j\rangle}~[(w^{\Sigma}_{ij}+\delta_{i\alpha}\delta_{j\alpha}\tau_\alpha^2
e^{-iq})~\Sigma_{i}^\dagger \Sigma_{j}+(w^T_{ij}+\delta_{i\alpha}\delta_{j\alpha}\tau_\alpha^2
e^{-iq})~\sum_{p=1}^{3}~T^{p\dagger}_{i}T^p_j]\nonumber\\
&=&H^{eff}_{\Sigma}+\sum_{p=1}^{3}H^{eff}_{T^p}.
\end{eqnarray}
We describe briefly the next steps. First, we employ the
Landauer-Buttiker formula to connect the transmission through the
dot with the retarded Green function:

\begin{equation}
T_{\sigma,\sigma'S,S'}(E)= 4 \tau_0^4 {sin^2}q
~|<\alpha,S|G^+_{\sigma \sigma'}(E)| \alpha',S'>|^2.
\end{equation}
The above formula assumes equal coupling strength to the leads
($\tau_{0}$) connected with the sites (states) $\alpha$, $\alpha'$.
"$S$" stands for the spin state of the magnetic impurity that is
changed to "$S'$" after the the scattering . The electron spin is
also changed from $\sigma$ to $\sigma '$ and only the transitions
that conserve the total spin give non-zero transmittances. The
parameter "$q$" is the impulse of the incident electron (see
\cite{Aldea} and references therein for details about the
dispersion formula in the leads and connection with the
Landauer-Buttiker formalism).

At this point we remember that, for a single electron case, the
retarded Green function is equal to the resolvent, and can be
calculated by:
\begin{equation}
G^+(E)=(E-H^{eff})^{-1}.
\end{equation}
The following definitions are introduced :
\begin{eqnarray}
G^+_{\Sigma}(E)=(E-H^{eff}_{\Sigma})^{-1},\nonumber\\
G^+_{T^p}(E)=(E-H^{eff}_{T^p})^{-1}.
\end{eqnarray}
All triplet Green functions are equal (we give up the index "$p$")
and the transmittance can be written:
\begin{equation}
T_{total}=\sum_{\sigma,\sigma'S,S'}T_{\sigma,\sigma'S,S'}(E)= 4
\tau_0^4 {sin^2}q
~(\frac14|G^+_{\Sigma}(E)|^2+\frac34|G^+_{T}(E)|^2).
\end{equation}
\begin{figure}[ht]
\centering
 \epsfxsize=0.5\textwidth \epsfbox{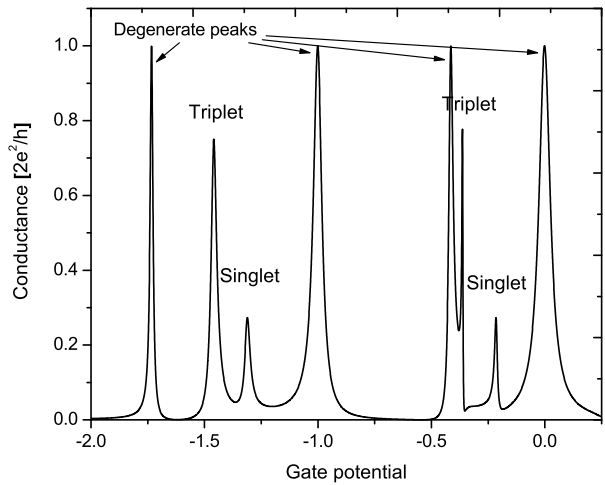}
  \caption {The conductance through a many-site quantum dot with a magnetic
  impurity, in the absence of electron-electron interaction (only the interaction with the magnetic impurity is considered)
  . Some peaks split into triplet and singlet,
  with the heights 3/4 and 1/4, respectively. The peaks corresponding to trajectories that avoid the impurity
  do not split and remain degenerate.}
\end{figure}
We show in Fig.1 the transmittance peaks for a quantum dot
modeled by a $3\times 5$ two-dimensional lattice. The exchange
interaction causes the singlet-triplet splitting of the
spectrum and the conductance peaks. The triplet peaks have the
hight $3/4$ and the singlet peaks $1/4$. For a realistic 2D
dot, certain eigenstates may be $zero$ at the impurity
position. Obviously, the corresponding transmittance peaks will
not  split. This case is also captured in Fig.1 (the
"degenerate peaks" do not undergo the singlet-triplet
splitting).

\section{Small quantum dots with magnetic impurities}
We saw in the previous section that the non-interacting dot is easy
to be modeled by a $n \times m$ lattice, but the problem becomes
(technically) much more difficult when we include the Coulomb
interaction. A general Hamiltonian, that includes the Coulomb
interaction between any two electrons, can be written: : $H=H_{S}+
\sum_{i,j,\sigma,\sigma'}U_{i,j}c_{i,\sigma}^\dagger
c_{i,\sigma}c_{j,\sigma'}^\dagger c_{j,\sigma'}$. There are several
ways to address such a complex Hamiltonian, that allows only
approximate solutions with considerable difficulty. In this section
, we shall restrict to the case of a single-site dot with Coulomb
interaction and exchange interaction with a magnetic impurity. In comparison
to the Hamiltonian Eq.1, we shall introduce the Coulomb
interaction, but also we shall write explicitly, from the beginning,
the Hamiltonian of the leads and the leads-dot hopping term. The reason is the following:
in the previous section we used the effective Hamiltonian trick, meaning that leads
are introduced by a selfenergy term in the coupling sites. But this is no longer possible
if we consider electronic correlations or the Kondo effect, when the role of the leads
is more complex.

We re-write the Hamiltonian, in a convenient notation:
\begin{eqnarray}
H&=&\sum_{k,\sigma,\alpha}{\epsilon_k
c^\dagger_{k\sigma,\alpha}c_{k\sigma,\alpha}}+\epsilon_0\sum_{\sigma}{c^\dagger_{0\sigma}c_{0\sigma}}+
Uc^\dagger_{0\uparrow}c_{0\uparrow}c^\dagger_{0\downarrow}c_{0\downarrow}+\nonumber\\
\frac12&J&(c^\dagger_{0\uparrow}
c_{0\uparrow}-c^\dagger_{0\downarrow} c_{0\downarrow})S^z +
\frac12J~(c^\dagger_{0\uparrow} c_{0\downarrow} S^-+c^\dagger_{0\downarrow} c_{0\uparrow} S^+
) +\sum_{k,\sigma,\alpha}{t_\alpha(c^\dagger_{0\sigma}c_{k\sigma,\alpha}+h.c.)}\;,
\end{eqnarray}
where the first term represents the electrons in the lead $\alpha=
L$, $R$. The second term stands for the electronic level
$\epsilon_0$ in the dot. The next four terms describe
interactions: the Coulomb interaction of electrons with the opposite
spin orientation at the level $\epsilon_0$ and exchange interactions
with the magnetic impurity.
The last term in the Hamiltonian (Eq.10) corresponds to the coupling
between the quantum dot and the leads.  It shall be
considered, for simplicity, $t_L=t_R=t$.

Now, we want to determine the conductance for the model described by
the Hamiltonian (Eq.10). The current can be expressed by means of the
non-equilibrium Green functions as \cite{Meir-Landauer, Haug} :
\begin{equation}
j = \frac{ 2e}{h}\Gamma
\int{d\omega[f_L(\omega)-f_R(\omega)](-\mathrm{Im}\langle\langle
c_{0\uparrow}|c_{0\uparrow}^\dagger\rangle\rangle),}
\end{equation}
where $\Gamma=2\pi t^2\rho$ and $\rho=1/2D$ is the DOS for the
square band approximation; the half-width $D$ will be taken as
unity. $f_L$ and $f_R$ are the Fermi distribution functions in the
left and right leads, respectively. $\langle\langle
c_{0\uparrow}|c_{0\uparrow}^\dagger\rangle\rangle$ is the retarded
single particle Green function for an electron with the spin
$\sigma=\uparrow$ at the QD, which can be determined by the equation
of motion (EOM).

\begin{figure}[ht]
\epsfxsize=0.4\textwidth \epsfbox{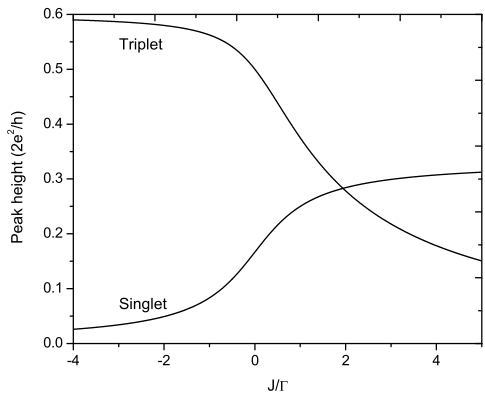}
  \caption {The height of the singlet and triplet peaks, that depend on the ratio $J/\Gamma$
  (exchange interaction versus coupling strength with the leads) , if
  the Coulomb interaction and electronic correlations are considered.
   }
  \end{figure}

The generic equation of motion (EOM) for the energy dependent retarded Green
function is given by
\begin{equation}
\omega\langle\langle A|B
\rangle\rangle=\langle\{A,B\}\rangle+\langle\langle
[A,H]|B\rangle\rangle\;,\end{equation} where $\langle\{A,B\}\rangle$
is the thermal average of the anticommutator between the operators A
and B. The procedure for solving the EOM for the retarded green
functions was described in detail in \cite{TB}. Basically, the
equations of motion introduce higher order Green functions, which result by performing the commutation
(with the Hamiltonian) required by the formula above. For a finite system, even in the presence of interactions, one can write
a finite number
of independent equations and the system closes (even if sometimes the number of
equations is very big, it is still solvable)
. But for our case, interacting system + infinite leads, one has an infinite set of equations,
and approximations are needed to close the system.

We propose the next approximation (that is valid at high temperatures \cite{bulka04})
\begin{equation}2t\sum_k\langle\langle
c_{k,\sigma}A|c^\dagger_{0,\sigma}\rangle\rangle\approx-i\Gamma\langle\langle
c_{0,\sigma}A|c^\dagger_{0,\sigma}\rangle\rangle .\end{equation}
 The
following solution is obtained for the conductance through the
singlet and triplet peaks (we take $U\rightarrow\infty$):

\begin{equation}
{\mathcal G}={\mathcal G}_S+{\mathcal G}_T
=\frac{1+2\phi_T}{2(3+4\phi_S-4\phi_S\phi_T)}R_S+\frac{3(1+2\phi_S)}
{2(3+4\phi_S-4\phi_S\phi_T)}R_T\;,
\end{equation}
where
$R_S=(2e^2/h)\times\Gamma^2/[\Gamma^2+(E_F-\epsilon_0+3J/4)^2]$ and
$R_T= (2e^2/h)\times\Gamma^2/[\Gamma^2+(E_F-\epsilon_0-J/4)^2] $
corresponds to the resonant conductance through the singlet and the
triplet level, respectively,
$\phi_S=\arctan[(\epsilon_0-E_F-3J/4)/\Gamma]/\pi$ and
$\phi_T=\arctan[(\epsilon_0-E_F+J/4)/\Gamma]/\pi$. The fractions
that multiply the resonances $R_{S(T)}$ contain the information
about the height of the conductance peaks. The ground and excited
states role changes with the sign of $J$, and the dependence on the
coupling also changes, being much more pronounced for the excited
state. All these effects are incorporated in the above formula.

By employing further the simplified form Eq.(14), one can estimate
the ratio between the heights of the conductance for the singlet and
triplet peaks. We notice that in Eq.(14), the parameter $\phi_S$
vanishes at the singlet resonance (found with the condition $\epsilon_0-E_F-3J/4=0$).
$\phi_T$ vanishes at the triplet resonance.
Therefore, one can straightforward calculate the ratio of the
conductance peaks
\begin{equation}
\frac{\max[\mathcal
{G}_{S}]}{\max[\mathcal{G}_{T}]}=\frac{(1+2\phi_0)(3-4\phi_0)}
{9(1-2\phi_0)}.
\end{equation}
where $\phi_0=\arctan(J/\Gamma)/\pi$.
 Fig.
2 gives a graphical image of the peaks height dependence on the
parameter $J/\Gamma$. For the antiferromagnetic
coupling the electronic transmission through the singlet state
dominates and in the limit  $J\gg \Gamma$ this ratio goes to
infinity and the triplet peak disappears. In the case of the
ferromagnetic coupling and $|J|\gg \Gamma$ we have opposite
situation - the singlet states disappears from transport and only
the peak corresponding to the triplet ground state is visible.

\begin{figure}[ht]
\epsfxsize=0.5\textwidth \epsfbox{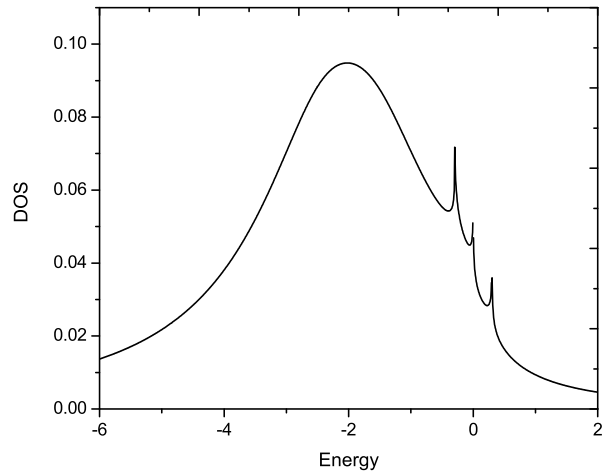}
  \caption {The density of states for a quantum dot with a magnetic impurity, in the Kondo regime.
  Notice the formation of three Kondo peaks, one at the Fermi energy ($E_F=0$) and two side peaks
  at a distance equal to the exchange interaction ($J=0.3$)  }
\end{figure}

In the last part of the paper, we give a very brief analysis of  the
Kondo regime for the impurity dot. It is known that a Kondo peak in
the density of states, at the Fermi energy, is present only if a
degenerate level exists below the Fermy energy. This would allow for
fluctuations (in the "classical" Kondo one has spin fluctuations),
that involve also the electrons from the leads which, for this reason,
can pass through the dot with increased probability.
In our case, such a
degenerate level would be the triplet and the system shows the central Kondo
peak (at the Fermi energy). This central peak is important for transport,
because at low temperature and low bias, practically only the electrons around the
Fermy level contribute to the conductance.

On the other hand, the singlet-triplet level
structure should
allow also the observation of side Kondo peaks, similar to the case of an
applied magnetic field \cite{Meir_magnetic}. We obtain a three-peaks
Kondo structure and the typical density of states is plotted in
Fig.3. The calculations were performed following the decoupling
proposed in \cite{Lacroix,Meir91} (see \cite{TB} for calculations details)
for temperatures closed to the Kondo temperature.

\section{Conclusions}
In this paper, we present two models of quantum dots with magnetic
impurities. The first model is a many-site lattice model (large
quantum dots). It neglects the Coulomb interaction and many-body
correlations, but can be used to model realistic geometries. In the
absence of the magnetic impurity, the transmittance shows peaks with
the height $1$ (perfect transmittance). If a magnetic impurity is
placed in the dot, one has triplet peaks with height $3/4$ and
singlet peaks with height $1/4$. Some eigenstates inside the dot may
avoid the impurity position, and the corresponding peaks do not
split into singlet and triplet.

The second model we propose is a single-site dot with Coulomb
and exchange interactions. Here we give up
the geometrical aspects, in order to focus on the correlations
effects. Such a model is realistic for small quantum dots, where the
Coulomb is strong and the level-spacing is large. The singlet and
triplet conductance peaks are shown to depend on the ratio
$J/\Gamma$ (the exchange interaction versus the coupling strength
with the leads). In particular, the transmission through the excited
states depends strongly on the coupling to the leads.

In the Kondo regime, the presence of the $S=1/2$ magnetic
impurity generates three Kondo peaks in the density of states.

\acknowledgments{The work was supported as a part of ESF EUROCORES
Programme FoNE by funds from Ministry of Science and Higher
Education (Poland) and EC 6FP (contract N. ERAS-CT-2003-980409), and
EC project RTNNANO (contract N. MRTN-CT-2003-504574). M.T and A.A.
acknowledge support from the Romanian excellence programme CEEX
D-11-45.}


\begin{thebibliography}{99}
\bibitem{explain}The sign convention and the $1/2$ factor ("$+\frac12J$")
are different from Ref. \cite{Aldea}, in order to use a unified notation throughout this paper.
\end{thebibliography}
\end{document}